\documentclass[%
 reprint,
 amsmath,amssymb,
 aps,prd
]{revtex4-2}

\usepackage{graphicx}
\usepackage{dcolumn}
\usepackage{bm}
\usepackage{makecell}
\usepackage[colorlinks=true,linkcolor=blue,citecolor=blue,urlcolor=blue]{hyperref}

\newcommand{\finesse}{\textsc{Finesse}}
\renewcommand{\Re}{\operatorname{Re}}
\renewcommand{\Im}{\operatorname{Im}}

\begin{document}


\preprint{APS/123-QED}

\title{Tuning of resonant doublets in coupled optical cavities}

\author{Riccardo Maggiore}
\affiliation{Department of Physics and Astronomy, VU Amsterdam; De Boelelaan 1081, 1081, HV, Amsterdam, The Netherlands}
\altaffiliation{Nikhef; Science Park 105, 1098, XG Amsterdam, The Netherlands}
\email{r.maggiore@nikhef.nl}

\author{Artemiy Dmitriev}
\affiliation{Institute of Gravitational Wave Astronomy, School of Physics and Astronomy, University of Birmingham, Birmingham B15 2TT, United Kingdom}
\email{admitriev@star.sr.bham.ac.uk}

\author{Andreas Freise}
\affiliation{Department of Physics and Astronomy, VU Amsterdam; De Boelelaan 1081, 1081, HV, Amsterdam, The Netherlands}
\altaffiliation{Nikhef; Science Park 105, 1098, XG Amsterdam, The Netherlands}
\email{a.freise@nikhef.nl}

\author{Mischa Sallé}
\affiliation{Department of Physics and Astronomy, VU Amsterdam; De Boelelaan 1081, 1081, HV, Amsterdam, The Netherlands}
\altaffiliation{Nikhef; Science Park 105, 1098, XG Amsterdam, The Netherlands}
\email{msalle@nikhef.nl}

\date{\today}


\begin{abstract}
The mode profile of a coupled optical cavity often exhibits a resonant doublet, which arises from the strong coupling between its sub-cavities. Traditional readout methods rely on setting fields of different frequencies to be resonant in either sub-cavity, which is challenging in the case of strong coupling. In this regime, the coupled cavity behaves as a single resonator, and a field must be resonant in all its parts. Consequently, specialized sensing schemes are necessary to control strongly coupled cavities. To address this issue, we propose a novel technique for the relative measurement of the degrees of freedom of a strongly coupled cavity. Our approach enables simultaneous frequency stabilization and fine-tuning of frequency splitting in the resonant doublet. Overall, our proposed technique offers a promising solution to control the properties of coupled cavities, facilitating advanced applications in the fields of gravitational-wave detection, quantum cavity optomechanics, and other related areas.
\end{abstract} 

\maketitle


\section{\label{section:1}Introduction}

Large scale gravitational-wave detectors are operational and the design of third-generation detectors is under way. Current gravitational-wave detectors, such as LIGO~\cite{Aasi_2015}, Virgo~\cite{Acernese_2015} and KAGRA~\cite{Somiya_2012} use dual-recycled Michelson interferometers with arm cavities. Next-generation detectors, such as the Einstein Telescope~\cite{Et2020} and Cosmic Explorer~\cite{Ce}, are expected to inherit the same design concept. Achieving the desired increase in sensitivity for these new detectors will be challenge and requires the development of new noise-mitigation strategies to overcome the limits of current laser interferometers. Many of these techniques rely on the ability to operate resonant optical systems under strict stability criteria. Theoretical design studies start to explore the use of strongly coupled optical cavities as key elements in new precision metrology designs. As an example, coupled cavities can be used in cavity optomechanics to enhance the displacement sensitivity, cooling, and effects relying on strong quantum back-action~\cite{dobrindt2010,aspelmeyer2014}. Furthermore, coupled cavities can be used to achieve stable behavior of phase-correcting quantum phase-insensitive filters~\cite{Li2020,dmitriev2022} and to explore PT-symmetric systems based on balanced gain and loss~\cite{jing2014,ozdemir2019}. 
More generally, developing the ability to control coupled cavities would allow us to explore a wider array of optical setups, including next-generation gravitational-wave detectors. For example, squeezed light is used in current detectors to increase quantum-noise limited sensitivity~\cite{PhysRevLett.110.181101}. Future detector designs, incorporating frequency-dependent squeezing, demand the addition of one or more filter cavities~\cite{PhysRevD.65.022002}. This cascade of filter cavities can potentially be replaced with a coupled cavity, which was found to perform identically for the ET-LF case~\cite{PhysRevD.101.082002}. 

\begin{figure*}[!t]
\centering
\begin{minipage}{0.475\textwidth}
  \centering
  \includegraphics{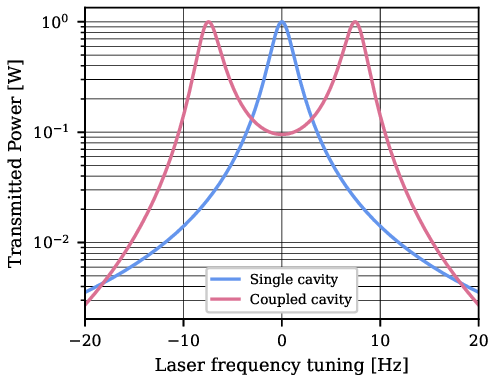}
\end{minipage}
\hfill
\begin{minipage}{0.475\textwidth}
  \centering
  \includegraphics[width=\linewidth]{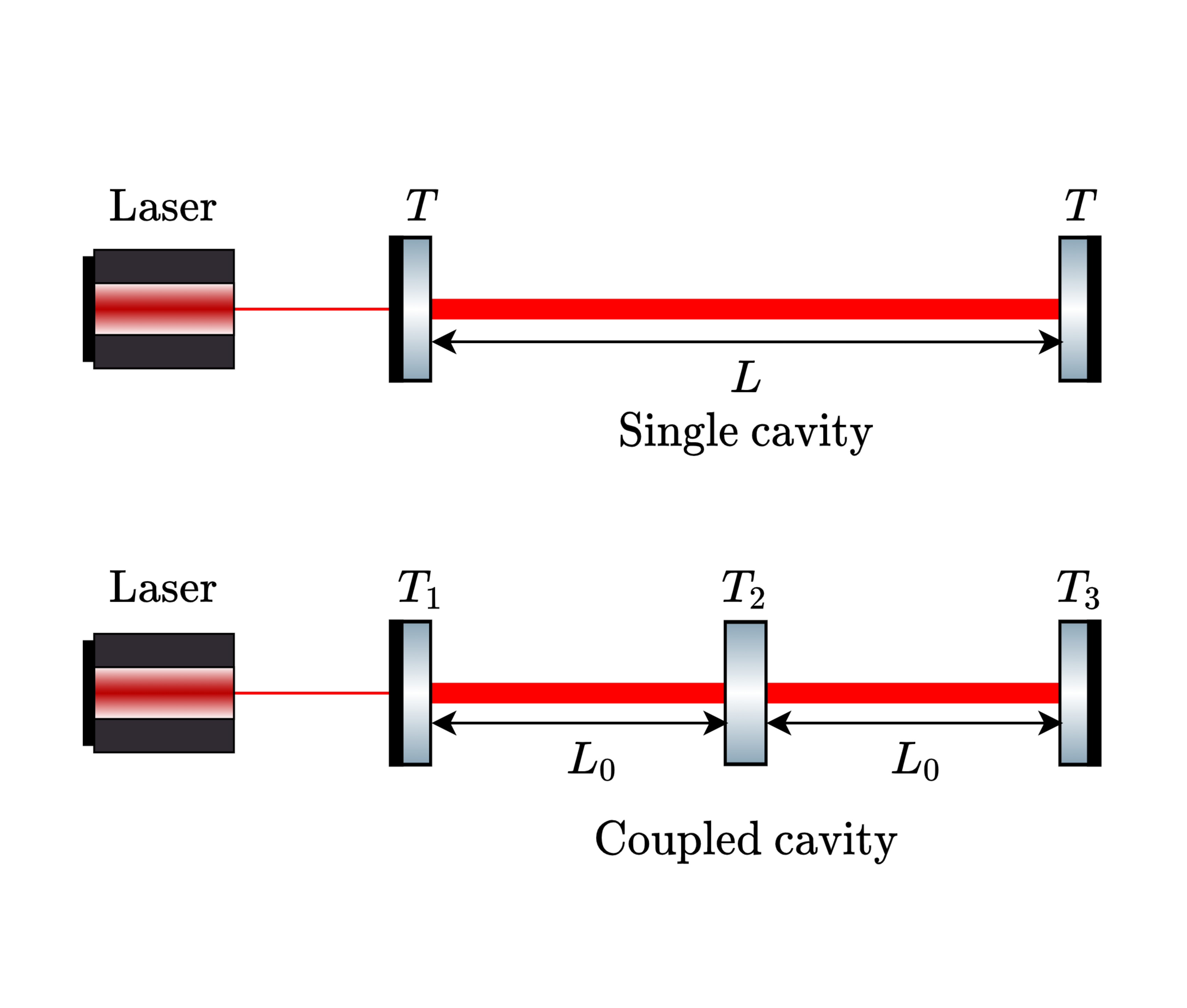}
\end{minipage}

\caption{Resonances in a single and a coupled cavity. A schematic of a single and coupled cavity set-up is on the right hand side. }
    \label{figure:1}
\end{figure*}

For strongly coupled cavities to be used in precision experiments, such as gravitational-wave detectors, we must develop robust schemes to sense and control parameters of the cavities with a sufficient shot-noise limited sensitivity. 
Controlling coupled cavities is not unprecedented, as demonstrated by gravitational-wave detectors. These detectors are typically comprised by several coupled optical resonators whose degrees of freedom are measured by well-established methods, as outlined in~\cite{Evans2002}. These sensing schemes are based on the possibility of setting optical fields of different frequency to be resonant in different parts of the detector~\cite{staley2014}.
However, these techniques are not applicable to strongly coupled cavities that exhibit resonance splitting. 

The mode profile of a single optical cavity is given by the Airy distribution~\cite{Ismail:16}. In the case of coupled cavities, if the coupling is strong enough, the mode profile turns from a single peak into a resonant doublet, which can be used to enhance multiple modes simultaneously and increase the power driving mitigation systems -- Fig.~\ref{figure:1}. In this regime, if a field is resonant in one of the sub-cavities, it is simultaneously resonant in the other one. The system behaves as single resonator and a field is either resonant in the coupled cavity or it is not. Thus the resonance condition of a single field is not sufficient to read out the parameters of the cavity. Therefore, controlling strongly coupled cavities necessitates the development of specialized sensing techniques.

This paper presents a novel approach to measuring a coupled cavity’s degrees of freedom under the condition of strong coupling between the sub-cavities. This technique enables the tunability of the splitting between the resonant frequencies, which might be crucial for many applications.

The outline of this paper is the following. In section~\ref{section:2}, we go through the basics of coupled cavities and show how the distance between the two resonances can be tuned. In section~\ref{section:3}, we explain the readout scheme, which consists of phase modulation in combination with a beat frequency demodulation scheme. Section~\ref{section:4} is dedicated to the examination of the fundamental noise limits of the readout scheme set by the quantum nature of light. Appendix~\ref{appendix:freq_spl} shows how the transmission spectrum of a symmetric coupled-cavity system depends on system parameters and degrees of freedom. In appendix~\ref{appendix:A}, it is demonstrated that a coupled cavity can be mathematically treated as two independent single cavities in the vicinity of the resonant doublet. This approximation is utilized in appendix~\ref{appendix:B} to derive the analytical formulation of the readout functions. Additionally, in the calculation, it is shown that the error signal designed for laser frequency stabilization exhibits the same symmetries as the Pound-Drever-Hall (PDH)~\cite{Drever1983, Black2001} signal. As a consequence, the error signal is insensitive to several noise sources, at least at a first-order level. Numerical modeling in this paper has been done using the frequency-domain modeling software \finesse~\cite{FINESSE, finesse3}.

\begin{figure*}[!t]
\centering
\begin{minipage}{0.475\textwidth}
  \centering
  \includegraphics{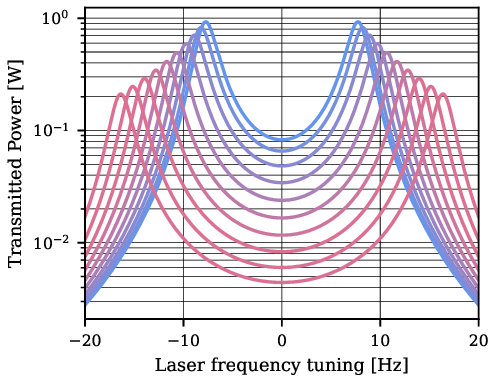}
\end{minipage}
\hfill
\begin{minipage}{0.475\textwidth}
  \centering
  \includegraphics[width=\linewidth]{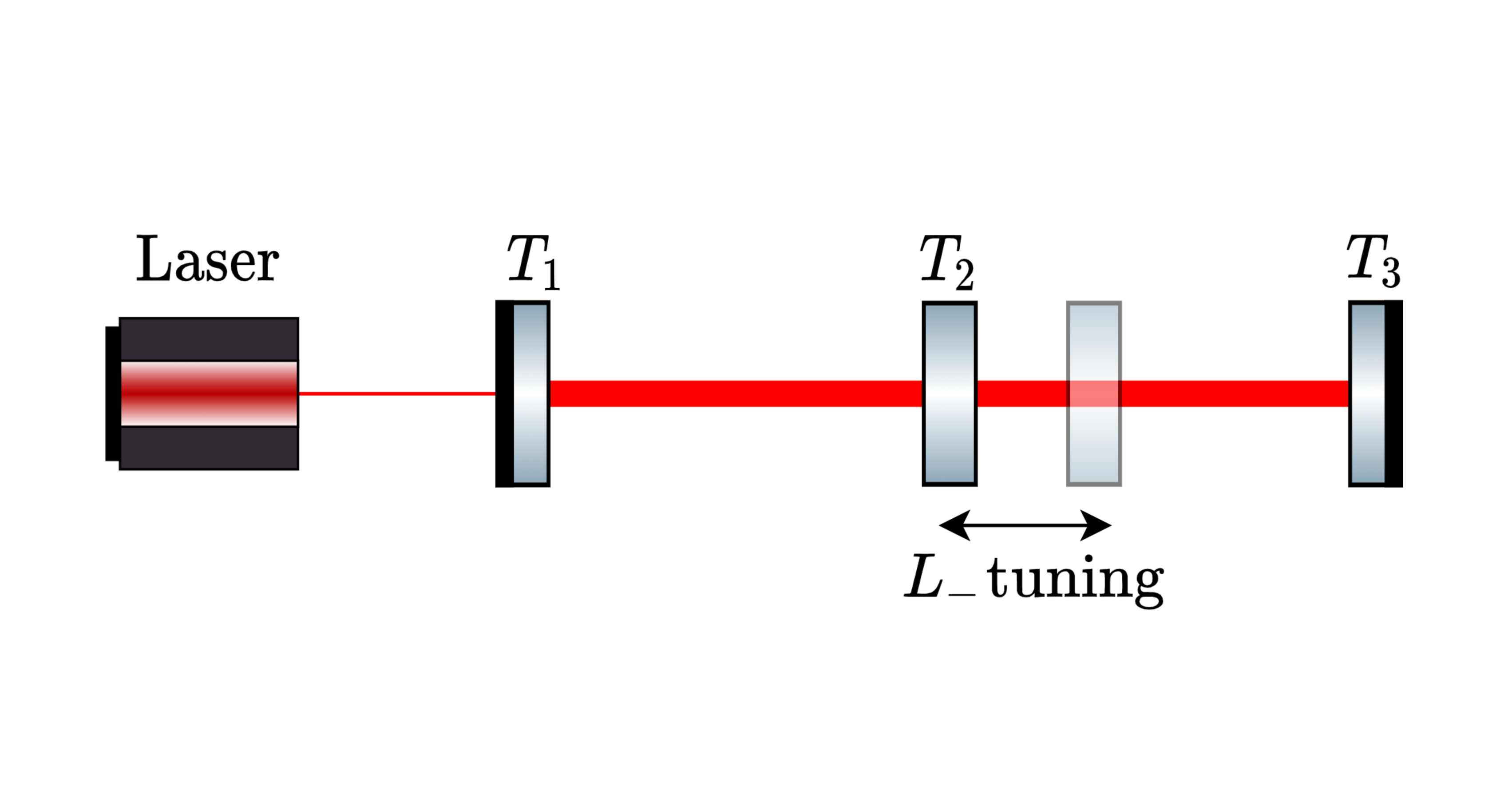}
\end{minipage}
\caption{Coupled-cavity resonant doublet profile. Blue is for $L_{-}=0$, red for increasing tunings of $L_{-}$. A schematic of a coupled cavity illustrating the tuning of $L_{-}$ is on the right hand side.}
    \label{figure:2}
\end{figure*}

\section{\label{section:2}System description}

A coupled cavity consists of two optically coupled resonators. We are consider a three-mirror setup, as illustrated in Fig.~\ref{figure:1}, with the corresponding parameters provided in Tab.~\ref{table:1}.

The mode profile of a coupled cavity exhibits a resonant doublet. This effect can be resolved if the bandwidth of the sub-cavities is smaller than the frequency splitting. If this is not the case, this coupling effect is hidden by the large bandwidth of the cavities and the mode profile of the system would consist of a single peak. The spacing between the resonances is dependent on the optical properties of the coupling mirror, as well as the length of the sub-cavities. We define the length $L_{1,2}$ of each sub-cavity as the sum of a macroscopic-scale length $L_0$ (which, for simplicity, we assume to be equal for both sub-cavities) and a microscopic-scale length variation $\delta L_{1,2}$. In our specific case, the macroscopic length is on the order of 10 km, while the microscopic lengths are in the range of 1 nm. Further details regarding the definition of these lengths can be found in appendix~\ref{appendix:freq_spl}. The frequency splitting between the resonances is influenced by both types of lengths, as follows:
\begin{equation}\label{eq:delta_f}
    \Delta f = \frac{c}{2\pi L_0}
    \arccos\left(\sqrt{R_2}\cos\frac{\omega_0 L_{-}}{c}\right).
\end{equation}
Here, $L_{-}$ is one of the two degrees of freedom (DoFs) of the coupled cavity, which we define as
\begin{subequations}\label{eq:DoFs}
\begin{align}
    L_{-} &= \delta L_{1} - \delta L_{2}, \\
    L_{+} &= \delta L_{1} + \delta L_{2}.
\end{align}
\end{subequations}
The derivation of Eq.~\ref{eq:delta_f} can be found in appendix~\ref{appendix:freq_spl}. It is worth noting that if $L_{-}=0$, the equation simplifies to
\begin{equation}\label{eq:delta_f_0}
    \Delta f_{0} = \frac{c\sqrt{T_2}}{2\pi L_0}.
\end{equation}
Here, $\Delta f_{0}$ is the minimum value of the frequency splitting that can be experienced by the coupled cavity. Tuning the differential cavities length $L_{-}$ increases the split from $\Delta f_{0}$ enabling fine-tuning to an arbitrary split $\Delta f$, as shown in Fig.~\ref{figure:2} and discussed in detail in Appendix~\ref{appendix:freq_spl}. This is a useful feature for any application that might require a precise tuning of the frequency splitting.

Tuning the total system's length $L_{+}$ is equivalent to shifting the frequency of the input laser beam.

\begin{table}[b!]
  \caption{Parameters for the example setup used in this work. The optics is assumed to be lossless. The order of magnitude of the coupled cavity parameters is taken from~\cite{PhysRevD.101.082002}.}
  \begin{ruledtabular}
    \begin{tabular}{lc}
      Parameter & Value \\
      \hline
      Laser power $P$ & $1$~W \\
      Modulation depth $\beta$ & $0.1$ \\
      Modulation frequency $f_{m}$ & $25$~Hz \\
      Input and end mirrors transmissivity $T_{1}=T_{3}$ & $1000$~ppm \\
      Middle mirror transmissivity $T_{2}$ & $10$~ppm \\
      Sub-cavities length $L_{0}$ & $10$~km \\
      Minimum frequency splitting $\Delta f_{0}$ & $15$~Hz \\
      Frequency splitting tuned to $\Delta f$ & $50$~Hz \\
    \end{tabular}
  \end{ruledtabular}
  \label{table:1}
\end{table}

\section{\label{section:3}Readout scheme}

\begin{figure}[!b]
    \centering
    \includegraphics[width=\columnwidth]{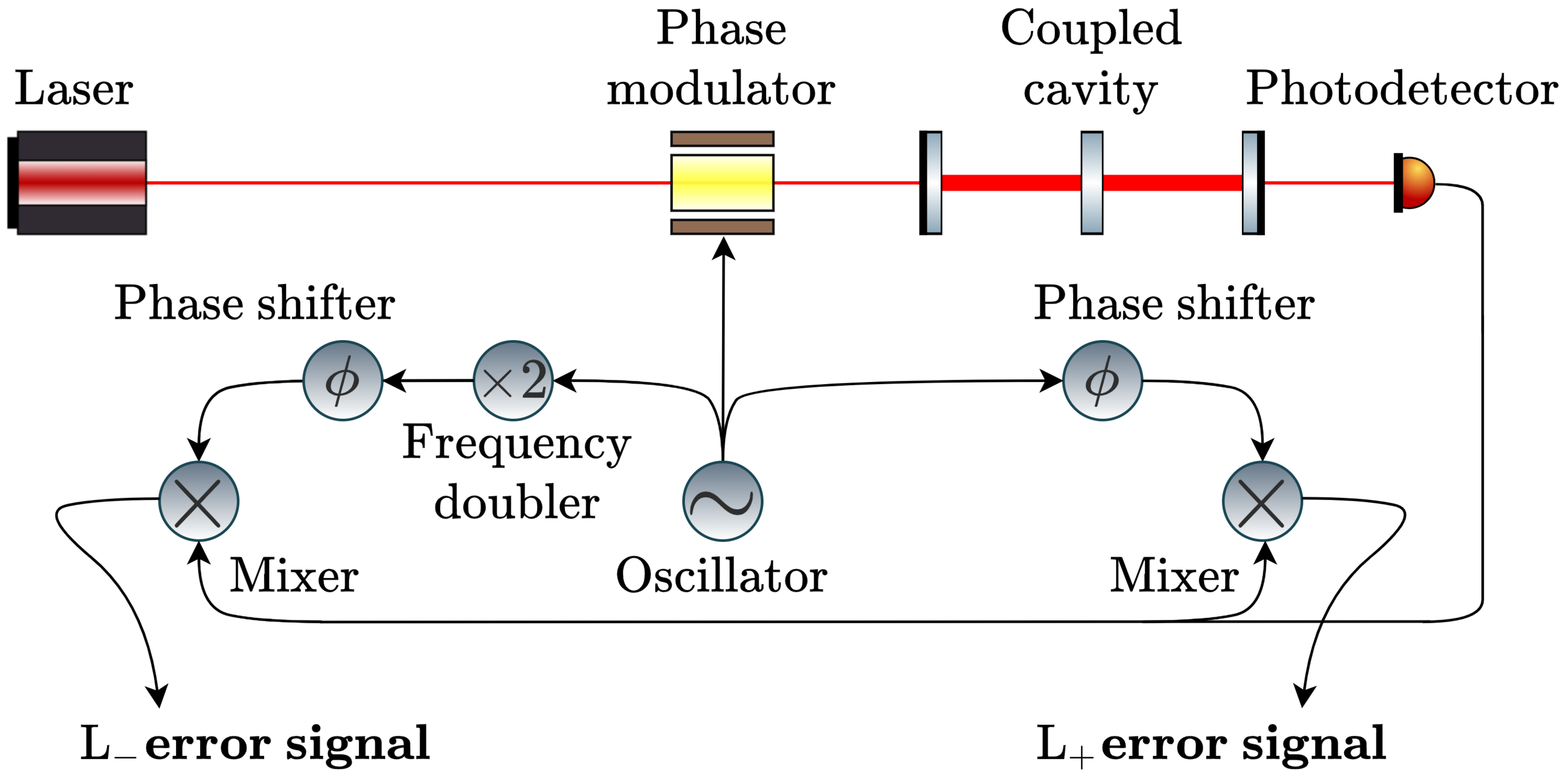}
    \caption{Schematic of the proposed readout scheme. A sinusoidal signal from the oscillator is used to drive the phase modulator, which generates sidebands on the laser light. The photodetector signal is then demodulated by mixing it with both the local oscillator signal and its second harmonic.}
    \label{figure:3}
\end{figure}

To derive a relative measure of the DoFs, we present a sensing scheme (Fig.~\ref{figure:3}) that allows for fine-tuning the frequency splitting between the resonances. This can be used as part of a feedback system to maintain the tuning of the coupled cavity.

\begin{figure*}[!t]
    \centering
    \begin{minipage}{0.45\textwidth}
        \includegraphics[width=0.95\linewidth]{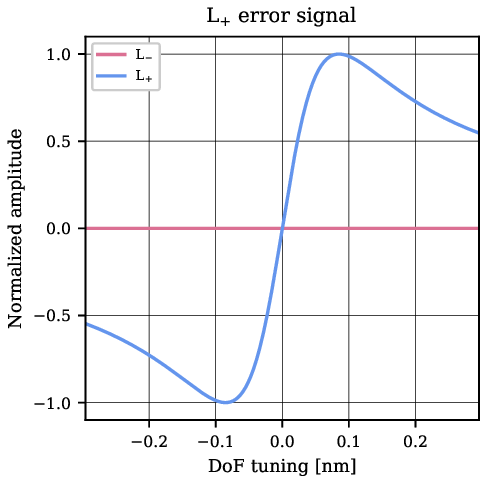}
        \includegraphics[width=0.95\linewidth]{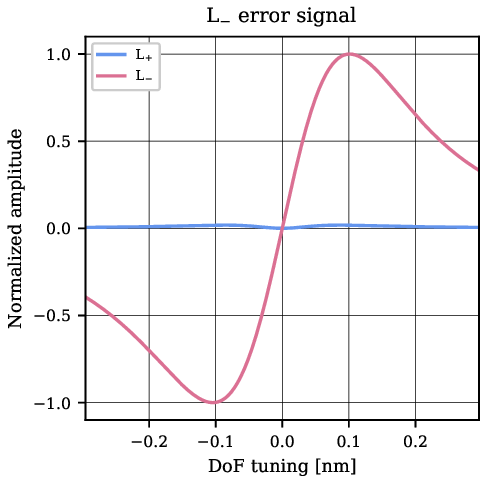}
    \end{minipage}
    \hfill
    \begin{minipage}{0.45\textwidth}
        \includegraphics[width=0.95\linewidth]{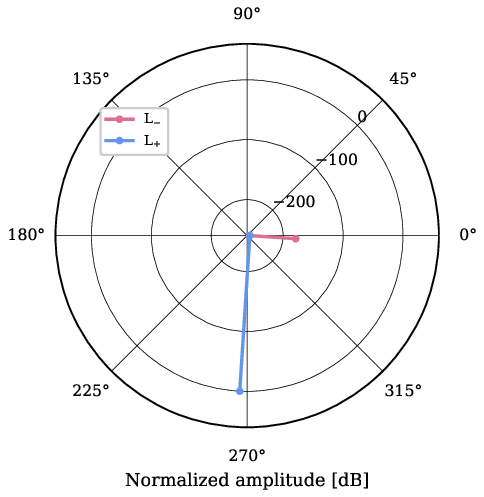}
        \includegraphics[width=0.95\linewidth]{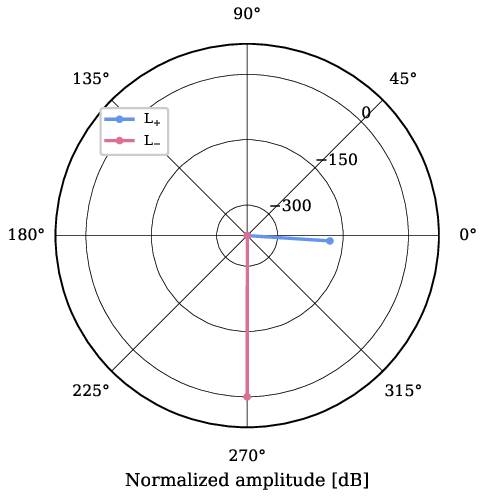}
    \end{minipage}
    \caption{The left column shows error signals plotted against the tuning of $L_{+}$ in blue and $L_{-}$ in red, with the tuning being relative to the operating conditions. In the right column, the compass plots show the amplitude and phase of the error signals' response to the motion of each DoF.}
    \label{figure:4}
\end{figure*}

The input optical field is phase modulated at half of the desired frequency splitting, $f_{m} = \Delta f/2$. The transmitted signal is mixed with a local oscillator signal of frequency $f_{m}$ and its second harmonic $2f_{m}$. Respectively, the error signals produced by demodulating at $f_{m}$ and $2f_{m}$ are sensitive to $L_{+}$ and $L_{-}$ motion, Fig.~\ref{figure:4}. In the same figure, the compass plots depict how each signal responds to the motion of each degree of freedom, showing that the degrees of freedom are well decoupled in sensing.

At the operating point, $L_{+}$ is adjusted to position the laser frequency at the center of the doublet, while $L_{-}$ is adjusted to ensure the frequency splitting corresponds to $2f_{m}$, resulting in resonance of both sidebands. If any application requires the sidebands not to be resonant, an alternative is to modulate the input laser at slightly different frequency than  $\Delta f/2$ and add an offset to $L_{-}$ error signal.

The error signal for $L_{-}$ is an odd function that indicates not only whether there is a mismatch between the output frequency $\Delta f$ and reference frequency $2f_{m}$, but also whether $\Delta f$ is greater or less than $2f_{m}$. This scheme would fail if $\Delta f=\Delta f_{0}$ as the splitting is at its minimum and it can only become larger. A positive or negative tuning of $L_{-}$ would increase the splitting to the same direction and readout function for $L_{-}$ becomes an even function unusable for control purposes.

In conclusion, it is important to note that demodulating a signal at $f_{m}$ not only isolates the signal at the demodulation frequency but also generates oscillating terms at its harmonics. These terms must be filtered using a low-pass filter, resulting in a control bandwidth of approximately $\sim 2f_{m}=\Delta f$.

Considering the chosen parameters, the control bandwidth is relatively narrow. However, these were deliberately selected to demonstrate the effectiveness of the sensing technique even in extreme scenarios, such as kilometer-scale cavities for a gravitational wave detectors. This effect must be considered in cavity design, but it does not restrict the technique that provides a more comprehensive modulation-demodulation approach for generating error signals in the case of strongly coupled cavities.

\section{\label{section:4}Shot-noise-limited resolution}
We examine in this section some fundamental noise limits of the error signals. Noise present in the $L_{+}$ error signal is indistinguishable from the laser's frequency noise, and similarly, any noise inherent in the $L_{-}$ error signal cannot be differentiated from the phase modulator's frequency noise. The quantum nature of light determines the ultimate limit of how low the noise level of an error signal can be.

At the operating conditions, the coupled cavity transmits only a small amount of power in the carrier, while the sidebands are completely transmitted. The average power reaching the photodetector is  $\sim 2P_{s}$, where $P_{s}$ is the power in each sideband at the output of the modulator. The corresponding shot noise in the power signal has an amplitude spectral density (ASD) of:
\begin{equation}
A = 2\sqrt{\frac{hc}{\lambda}P_{s}}.
\end{equation}
To estimate the frequency noise level in either signal, we divide the shot noise spectrum by the optical gain of each signal $D_{\pm}$. The derivation of the optical gains is available in appendix~\ref{appendix:B}. The resulting ASD for $L_{+}$ error signal is
\begin{equation}
    A_{+}=\frac{A}{D_{+}} =  \frac{\sqrt{T_2}}{16\pi L} \sqrt{\frac{hc^3}{\lambda \;P}}.
\end{equation}
For $L_{-}$ error signal, the ASD is
\begin{equation}
    A_{-}=\frac{A}{D_{-}} = \frac{T_1}{8\pi L}\frac{1}{\beta}  \sqrt{\frac{hc^{3}}{\lambda\; P}},
\end{equation}
where we have assumed the modulation depth $\beta$ to be small so to approximate the power of the sidebands as in~\cite{Bond2017}:
\begin{equation}
    P_{s} \simeq \frac{\beta^{2}}{4}P.
\end{equation}
The error signal for $L_{-}$ is obtained from the beating of one sideband with the other. Therefore, the optical gain of this signal is proportional to the power of the sidebands, as shown in appendix~\ref{appendix:B}. The lower the modulation depth, the lower the power of the sidebands, and as the previous equation indicates, the higher the shot noise.

By using feedback control, it is impossible to stabilize the frequency of both the laser and modulator beyond these limits\footnote{As referenced in appendix~\ref{appendix:freq_spl} and evident in Fig.~\ref{figure:2}, adjusting $L_{-}$ leads to a decrease in the power transmitted by the system, causing an increase in the shot noise of both signals. This effect is not considered in the equations, but it is relatively straightforward to incorporate. For example, if the average power received by the photodetector is reduced by a factor of $\gamma$, the optical gain of the error signals decreases by the same factor. As a result, the shot noise of both error signals increases by a factor of $\sqrt{\gamma}$.}.

As introduced in section~\ref{section:1} and further elaborated in appendix~\ref{appendix:B}, this scheme utilizes symmetries equivalent to the PDH scheme to generate the $L_{+}$ error signal. For the sake of comparison, the shot noise in a PDH signal is
\begin{equation}
    A_{\,\textnormal{PDH}}= \frac{T}{16\pi L} \sqrt{\frac{hc^3}{\lambda \;P}}.
\end{equation}
Consequently, the $L_{+}$ error signal would exhibit an equivalent noise floor as the PDH scheme would have for a symmetric cavity, where the transmission through each mirror is $T=\sqrt{T_2}$. The ratio between the ASDs of the two schemes can be expressed as:
\begin{equation}
    \frac{A_{+}}{A_{\,\textnormal{PDH}}} = \frac{\sqrt{T_2}}{T}.
\end{equation}

\section{Summary}

Controlling coupled cavities can enable exploration of a broader range of optical configurations. 
We show that a coupled cavity can be treated mathematically as two separate single cavities near the resonant doublet. 
This approximation is used to derive analytic expressions for the error signals. The analysis also reveals that the error signal created for laser frequency stabilization demonstrates the same symmetries as the PDH signal, which is convenient in terms of noise couplings.

We have illustrated how to generate a pair of error signals that, when combined with a feedback system, can maintain the tuning of a strongly coupled cavity. This technique utilises phase modulation and a beat-frequency demodulation scheme to anchor the mid-frequency of the doublet to the laser frequency and the spacing between the resonances to twice the modulation frequency. This enables the tuning of the degree of splitting between the resonances. Moreover, we have found that the degrees of freedom are well decoupled in sensing.

\appendix
\section{\label{appendix:freq_spl}The frequency splitting}
In this section, we show how the frequency splitting between resonant peaks in coupled cavities (Fig.~\ref{figure:2}) depends on cavity parameters. We consider the case of a three-mirror coupled cavity with equal transmission coefficients through the external mirrors, $T_{1}=T_{3}$, as well as the reflectivities $R_{1}=R_{3}=1-T_{1}$. The middle mirror's transmissivity is given by $R_2=1-T_2$. For simplicity, we also assume that the macroscopic length of both sub-cavities $L_0$ is the same, so that the sub-cavity lengths are
\begin{subequations}
\begin{align}
    L_1 &= L_0 + \delta L_1, \\
    L_2 &= L_0 + \delta L_2.
\end{align}
\end{subequations}
where the length deviations $\delta L_{1,2} \ll L_0$ are typically of the order of or smaller than a single wavelength of the EM field. Finally, we introduce the relative frequency of the electromagnetic (EM) field $\omega$ as a frequency offset between the absolute frequency of the EM field $\tilde\omega = \omega_0+\omega$ and one of the resonant frequencies $\omega_0\gg\omega$ of a single two-mirror cavity of length $L_0$ (i.e., so that $\omega_0 L_0 = \pi N$, where $N$ is an integer).

The transmission coefficient of the coupled cavity system is:
\begin{equation}\label{eq:coupled_cav}
    \xi(\omega) = \frac
    {
        -i T_1\sqrt{T_{2}}Z_1(\omega)Z_2(\omega)
    }
    {
        1-\sqrt{R_1 R_2}\left[Z_1^2(\omega)+Z_2^2(\omega)\right]+R_1 Z_1^2(\omega)Z_2^2(\omega)
    }.
\end{equation}
Functions $Z_1$ and $Z_2$ represent the extra phase gained by light as it propagates through each of the respective sub-cavities:
\begin{subequations}\label{eq:propagators}
\begin{align}
    Z_1(\omega) &= \exp\left( -i\tilde\omega L_1/c\right) \nonumber\\
    &\approx \exp\left( -i\omega \tau - i\omega_0 \delta L_1/c \right), \\
    Z_2(\omega) &= \exp\left( -i\tilde\omega L_2/c\right) \nonumber\\
    &\approx \exp\left( -i\omega \tau - i\omega_0 \delta L_2/c \right).
\end{align}
\end{subequations}
where $\tau=L_0/c$ represents the light's travel time across each of the sub-cavities. By switching to the degrees of freedom (DoFs) of interest (\ref{eq:DoFs}),
\begin{subequations}\label{eq:DoFs_inversed}
\begin{align}
    \delta L_1 &= (L_{+} + L_{-}) / 2 , \\
    \delta L_2 &= (L_{+} - L_{-}) / 2 ,
\end{align}
\end{subequations}
we can see that effect caused by any change in the common DoF $L_{+}$ is equivalent to a shift of the laser frequency $\omega$ by $\omega_0 L_{+}/(2 L_0)$. Therefore, the drift of $L_+$ would simply shift the whole resonance structure in the frequency domain without affecting the splitting between the resonant peaks.

By plugging equations (\ref{eq:propagators},~\ref{eq:DoFs_inversed}) into (\ref{eq:coupled_cav}) and assuming $L_{+}=0$, we can write the transmitted power as
\begin{widetext}
\begin{equation}\label{eq:coupled_cav_pow}
    \left|\xi(\omega)\right|^2 = 
    \frac{
        T_1^2 \, T_2
    }
    {
        \left(-2\sqrt{R_1R_2}\cos\varphi_{-}+(1+R_1)\cos(2\omega\tau)\right)^2
        +(1-R_1)^2\sin^2(2\omega\tau)
    },
\end{equation}
\end{widetext}
where
\begin{equation}
    \varphi_{-} = \frac{\omega_0 L_{-}}{c}
\end{equation}
is the additional phase shift caused by the differential DoF $L_{-}$. We notice that any non-zero $L_{-}$ would attenuate the transmitted power. By finding the roots of the derivative of function (\ref{eq:coupled_cav_pow}) and calculating the difference between the neighbouring ones, we express the frequency splitting of the resonance doublet as
\begin{equation}
    \Delta f = \frac{c}{2\pi L_0}
    \arccos\left(\frac{(1+R_1)}{2R_1}\sqrt{R_2}\cos\frac{\omega_0 L_{-}}{c}\right),
\end{equation}
or, if the peaks are well-resolved,
\begin{equation}\label{eq:delta_f_vs_Lminus}
    \Delta f \approx \frac{c}{2\pi L_0}
    \arccos\left(\sqrt{R_2}\cos\frac{\omega_0 L_{-}}{c}\right).
\end{equation}
If $L_{-}=0$, then the frequency splitting
\begin{multline}
    \Delta f = \Delta f_0 = \frac{c}{2\pi L_0} \arccos\sqrt{R_2} \\
    = \frac{c}{2\pi L_0} \arcsin\sqrt{T_2} \approx \frac{c \sqrt{T_2}}{2\pi L_0}
\end{multline}
is reduced to (\ref{eq:delta_f_0}). An extra factor of 2 that appears in these equations as compared to Equation 1 in~\cite{PhysRevD.99.102004} is due to the latter describing the frequency position of each of the split resonances with respect to the middle point rather than the full frequency splitting between them. The dependency of $\Delta f$ on the differential DoF $L_{-}$ is described by (\ref{eq:delta_f_vs_Lminus}).

\section{\label{appendix:A}Single cavities approximation}

In this section, we demonstrate that near the resonances, the transmission coefficient of a coupled cavity can be expressed as a linear combination of the transmission coefficients of two individual cavities. This is a powerful approximation as it allows to treat a coupled cavities as two single independent cavities.

We combine the transmission coefficients of two identical single cavities as follows:
\begin{equation}\label{eq:single_cav}
\begin{split}
  \tilde{\xi}(\omega) &= \frac{T\;e^{-i(\omega-\Omega)\tau}}{R\;e^{-2i(\omega-\Omega)\tau}-1} - \frac{T\;e^{-i(\omega+\Omega)\tau}}{R\;e^{-2i(\omega+\Omega)\tau}-1}.
\end{split}
\end{equation}
where $\Omega$ is defined on the frequency splitting of the coupled cavity as $\Omega=2\pi(\Delta f /2)$.

We assume that the travel time is equivalent to that in eq.~\ref{eq:coupled_cav}. As a consequence, the transmission coefficients in (\ref{eq:single_cav}) and (\ref{eq:coupled_cav}) will have the same free spectral range (FSR): $\Delta_{\,\textnormal{FSR}}=1/2\tau$. In the next steps, we will demonstrate that, near the resonances, $\xi(\omega)=\tilde{\xi}(\omega)$ provided that the reflectivity of the individual cavities is $R=\sqrt{R_{1}}$. After some algebra, the previous equation can be written as
\begin{equation}
    \tilde{\xi}(\omega) = \frac{-2i T\;\sin(\Omega\tau)\;e^{-2i\omega\tau}(e^{i\omega\tau}+R\;e^{-i\omega\tau})}{1-R\;[2\cos(2\Omega\tau)\;e^{-2i\omega\tau}-R\;e^{-4i\omega\tau}]}.
\end{equation}
For frequencies much much smaller that the FSR of the cavities, the transmission coefficient in the previous equation behaves as
\begin{equation}
     \tilde{\xi}(\omega) \simeq \frac{-2i T\;\sin(\Omega\tau)\;e^{-2i\omega\tau}(1+R)}{1-R\;[2\cos(2\Omega\tau)\;e^{-2i\omega\tau}-R\;e^{-4i\omega\tau}]}.
\end{equation}
Assuming high finesse for the equivalent single cavities, which implies $(1+R)\sim2$, the transmission coefficient $\tilde{\xi}(\omega)$ becomes\begin{equation}\label{eq:single_cav2}
    \tilde{\xi}(\omega) \simeq \frac{-4i T\;\sin(\Omega\tau)\;e^{-2i\omega\tau}}{1-R\;[2\cos(2\Omega\tau)\;e^{-2i\omega\tau}-R\;e^{-4i\omega\tau}]}.
\end{equation}
If the sub-cavities of a coupled cavity have the same length $L=L_{1}=L_{2}$, the frequency split is:
\begin{equation}
   \Delta f =  \frac{c\sqrt{T_{2}}}{2\pi L} = \frac{\sqrt{T_{2}}}{2\pi\tau}.
\end{equation}
The expression above can be used to re-write the cosine and sine terms in (\ref{eq:single_cav2}) as:
\begin{subequations}\label{eq:single_cav_apprx2}
\begin{align}
  \sin(\Omega\tau) &= \sin\left(\sqrt{T_{2}}/2\right) \simeq \sqrt{T_{2}}/2, \\
  \cos\left(2\Omega\tau\right) &\simeq \sqrt{1-(2\Omega\tau)^{2}} = \sqrt{R_{2}}.
\end{align}
\end{subequations}
Substituting (\ref{eq:single_cav_apprx2}) back into (\ref{eq:single_cav2}), we obtain 
\begin{equation}\label{eq:single_cav_final}
    \tilde{\xi}(\omega) \simeq \frac{-2i T\sqrt{T_{2}}\;e^{-2i\omega\tau}}{1-R\;[\sqrt{R_{2}}\;e^{-2i\omega\tau}-R\;e^{-4i\omega\tau}]}.
\end{equation}
As mentioned above, we assume
\begin{equation}
    R=\sqrt{R_{1}}.
\end{equation}
Under this assumption,
\begin{equation}
    T=1-\sqrt{R_{1}}\simeq (1-R_{1})/2=T_{1}/2,
\end{equation}
which makes (\ref{eq:coupled_cav}) and (\ref{eq:single_cav_final}) identical.

\begin{figure}
    \centering
    \includegraphics{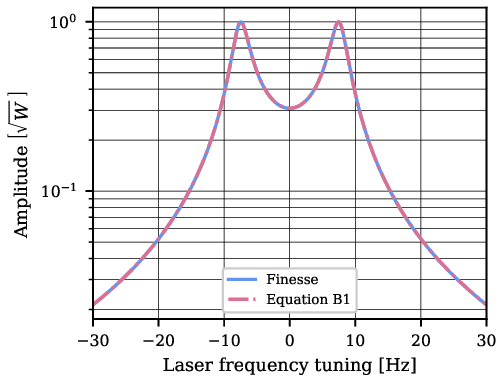}
    \caption{Comparison between the transmission coefficient of the coupled cavity computed using \finesse~against the approximation in eq.~\ref{eq:single_cav}, where we set $R = \sqrt{R_{1}}$.}
    \label{figure:5}
\end{figure}
A comparison between \finesse~output and the approximation (\ref{eq:single_cav}) can be found in Fig.~\ref{figure:5}.
This approximation reveals that a coupled cavity can be represented as an equivalent combination of individual symmetric cavities, with each mirror exhibiting a transmission of $T = T_1/2$. With this information, we can proceed to define several quantities that will prove valuable in the following analysis. To begin, we define the Finesse of the coupled cavity as:
\begin{equation}
\mathcal{F} = \frac{\pi}{T} = \frac{2\pi}{T_1}
\end{equation}
where we have implicitly assumed a high finesse. The linewidth of one of the resonant peaks is therefore given by:
\begin{equation}
    \Delta\nu_{\,\textnormal{LW}} = \frac{\Delta\nu_{\,\textnormal{FSR}}}{\mathcal{F}}
\end{equation}
where $\Delta\nu_{\,\textnormal{FSR}}$ has been defined above, but we repeat the definition for clarity:
\begin{equation}
    \Delta\nu_{\,\textnormal{FSR}} = \frac{c}{2L}
\end{equation}
Here, $L$ represents the length of one of the sub-cavities.

\section{\label{appendix:B}The error signals}

This section provides a mathematical interpretation of section~\ref{section:3} by assuming the same setup and deriving the analytical equations for the $L_{+}$ and $L_{-}$ error signals near the resonant doublet. We show that the $L_{+}$ error signal shares the same symmetries as the PDH error signal.

Following appendix~\ref{appendix:A}, we approximate the transmission coefficient of a coupled cavity to:
\begin{equation}\label{eq:single_cav_gamma}
  \xi(\omega) = \frac{T\;e^{-i(\omega-\Omega)\tau}}{R\;e^{-2i(\omega-\Omega)\tau}-1} - \frac{T\;e^{-i(\omega+\Omega)\tau}}{R\;e^{-2i(\omega+\Omega)\tau}-1},
\end{equation}
with
\begin{align*}
  R &= \sqrt{R_{1}}, \\
  T &= 1-R.
\end{align*}

\paragraph{Input beam modulation.}
The input carrier field is phase-modulated at $\Omega$ so three different beams are incident on the coupled cavity: a carrier of angular frequency $\omega_{c}$, and two sidebands, with angular frequencies $\omega_{\pm} = \omega_{c}\pm\Omega$. For small modulation depths $\beta$, the incident field can be expanded using Bessel functions as:
\begin{equation}
  E_{i} \simeq E_{0}\;\{J_{0}(\beta)e^{i\omega_{c} t} \;+\; J_{1}(\beta)[e^{i\omega_{+}t}\; - e^{i\omega_{-}t}]\},
\end{equation}
where $E_{0}$ is the amplitude the incident beam. The power in the carrier after modulation is
\begin{equation*}
    P_{c} = J_{0}^{2}(\beta)|E_{0}|^{2}
\end{equation*}
and the power in each of the first-order sidebands is
\begin{equation*}
    P_{s} = J_{1}^{2}(\beta)|E_{0}|^{2}.
\end{equation*}

\paragraph{The transmitted beam.}
To calculate the beam transmitted by the coupled cavity, we treat each frequency independently and multiply each field component by the transmission coefficient at the corresponding frequency:
\begin{multline}
E_{t} = E_{0}\,\xi(\omega_{c})J_{0}(\beta)e^{i\omega_{c}t} \\
\quad + E_{0}\,J_{1}(\beta)[\xi(\omega_{+})e^{i\omega_{+} t} - \xi(\omega_{-})e^{i\omega_{-}t}]
\end{multline}

The total power transmitted by the coupled cavity and measured by the photodetector is:
\begin{multline}\label{eq:analytics_Ptot}
  P_{t} = |E_{t}|^{2} \\
  = P_{c}\;|\xi(\omega_{c})|^{2} + P_{s}\;\{|\xi(\omega_{+})|^{2} + |\xi(\omega_{-})|^{2}\} \\
  \quad +2\sqrt{P_{c}P_{s}}\;\{\Re[K]\cos(\Omega t)+\Im[K]\sin(\Omega t)\} \\
  \quad -2P_{s}\;\{\Re[G]\cos( 2\Omega t) + \Im[G]\sin(2\Omega t)\} 
\end{multline}
with
\begin{align*}
  K &= \xi(\omega_{c})\xi^{*}(\omega_{+}) - \xi^{*}(\omega_{c})\xi(\omega_{-}), \\
  G &= \xi(\omega_{+})\xi^{*}(\omega_{-}).
\end{align*}

\begin{figure}[!t]
    \centering
    \includegraphics{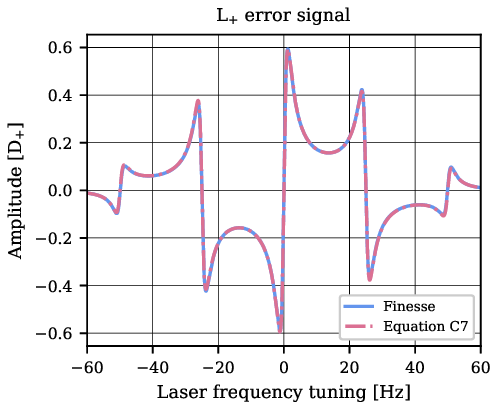}
    \caption{The $L_{+}$ error signal is plotted against the tuning of the laser frequency, as calculated using \finesse. The error signal obtained from eq.~\ref{eq:ES1_full} is also displayed in red.}
    \label{figure:7}
\end{figure}

Eq.~\ref{eq:analytics_Ptot} contains terms of angular frequency $\Omega$ that result from the beat between the carrier and the sidebands. By demodulating $P_{t}$ at $\Omega$, these terms can be isolated to produce the $L_{+}$ error signal. On the other hand, the $2\Omega$ terms arise from the beat between the upper and lower sidebands, and demodulating at $2\Omega$ generates the $L_{-}$ error signal.

The process of demodulation involves multiplying the output of the photodetector $P_{t}$ by a cosine function: $\cos(\tilde{\Omega}t + \theta)$. This cosine function acts as a local oscillator, with $\tilde{\Omega}$ representing the demodulation frequency and $\theta$ representing the demodulation phase. The product of this multiplication is then integrated over one period of the cosine function, effectively isolating the $\tilde{\Omega}$ terms of $P_{t}$.

In the following paragraphs, $P_{t}$ is demodulated at $\Omega$ and $2\Omega$ to obtain $L_{+}$ and $L_{-}$ error signals. Regarding the DoFs, they were initially defined in the main section in terms of lengths. However, as shown in Appendix~\ref{appendix:freq_spl}, they could also be defined in terms of frequencies. In particular, a change in $L_{+}$ is equivalent to a variation in the laser frequency, while a change in $L_{-}$ is equivalent to a variation in the modulation frequency. The following analysis presents the error signal in response to variations in both the laser frequency and modulation frequency.

\begin{figure*}
    \centering
    \begin{minipage}{0.45\textwidth}
        \includegraphics[width=0.975\linewidth]{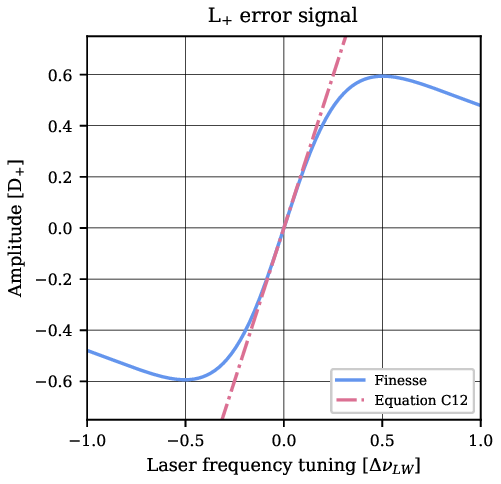}
    \end{minipage}
    \hfill
    \begin{minipage}{0.45\textwidth}
        \includegraphics[width=0.975\linewidth]{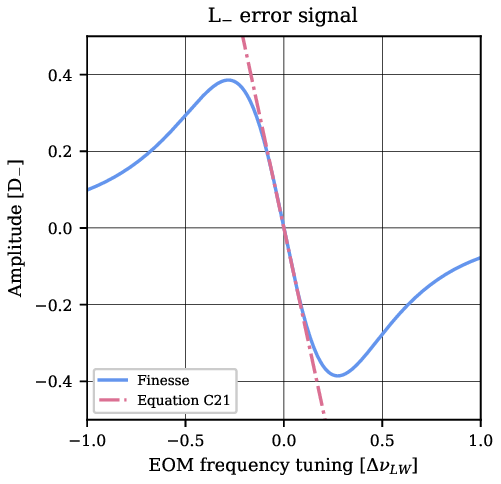}
    \end{minipage}
    \caption{Comparison between the behavior of the $L_{+}$ and $L_{-}$ error signals as computed using \finesse, and the eqs.~\ref{eq:ES1_lin} and~\ref{eq:ES2_lin}, which describe their behavior within their linear region. For these error signals, the frequency splitting has been tuned to $\Delta f = 30$~Hz. As explained in the main text, this leads to a reduction in transmitted power, resulting in a decrease in optical gain. This effect has been disregarded in the equations, while it is considered by default in the \finesse~model. In order to compare the two results, the \finesse~one has been multiplied by the attenuation of the transmitted power, which is $\gamma = 11$ for the given frequency splitting.}
    \label{figure:6}
\end{figure*}

\paragraph{Demodulating at $\Omega$: $L_{+}$ error signal.}
Assuming demodulation at $\Omega$, we isolate the $\Omega$ terms in eq.~\ref{eq:analytics_Ptot} to obtain $L_{+}$ error signal:
\begin{equation}\label{eq:ES1_IQ}
    \epsilon_{+} = 2\sqrt{P_{c}P_{s}}\;\left\{\Re[K]\cos(\theta)+\Im[K]\sin(\theta)\right\}.
\end{equation}
The input beam is modulated at $\Omega = 2\pi(\Delta f/2)$, thus the first-order sidebands are fully transmitted by the coupled cavity, $\xi(\omega_{\pm})\sim \mp 1$. We can write $K$ as:
\begin{equation}\label{eq:K}
    K = \xi(\omega_{c})\xi^{*}(\omega_{+}) - \xi^{*}(\omega_{c})\xi(\omega_{-}) \sim -2\;\Re\{\xi(\omega_{c})\},
\end{equation}
which is purely real. The sine term in eq.~\ref{eq:ES1_IQ} is negligible and we assume $\theta=0$. $L_{+}$ error signal is given as:
\begin{equation}\label{eq:ES1_full}
    \epsilon_{+} = 2\sqrt{P_{c}P_{s}}\;\Re\{K\} \sim  -4\sqrt{P_{c}P_{s}}\;\Re\{\xi(\omega_{c})\}.
\end{equation}

The above equation provides the analytical expression for the error signal we were seeking, and the comparison with the output of \finesse~is shown in Fig.~\ref{figure:7}. In the following lines, we will derive the equation that characterizes the error signal's linear behavior near the operating point.

The readout scheme sets the frequency of the carrier in the middle of the resonant doublet. In other words, the frequency of the carrier has to be an integer multiple of the free spectral range of the cavity:
\begin{equation}
    \omega_{c} = 2\pi N \cdot \Delta\nu_{\,\textnormal{FSR}},
\end{equation}
where $N$ is an integer. As we are interested in the behavior of the error signal around $\omega = \omega_{c}$, we use the Taylor expansion of $\xi{(\omega)}$ up to the first order around that point:
\begin{multline}\label{eq:ES1_taylor}
  \xi(\omega_{c}) \simeq 2i \; \Im \left[\frac{Te^{i\Omega\tau}}{Re^{2i\Omega\tau}-1}\right] \\
  \quad - \Im \left[\frac{T \; e^{i\Omega\tau}\;(1+Re^{2i\Omega\tau})}{(Re^{2i\Omega\tau}-1)^{2}} \right] \frac{\delta\omega_{c}}{\Delta\nu_{\,\textnormal{FSR}}},
\end{multline}
where $\delta\omega_{c}$ is the deviation of the laser frequency from the center of the doublet. In this scheme, the modulation frequency is much smaller than the cavity FSR, thus $\Omega\tau << 1$. This implies that the first term in brackets is almost purely real and therefore its imaginary part is null and $\xi(\omega_{c}) \sim 0$. By substituting eq.~\ref{eq:ES1_taylor} back into eq.~\ref{eq:K} and~\ref{eq:ES1_full}, $L_{+}$ error signal is given as:
\begin{equation}\label{eq:ES1_lin1}
    \epsilon_{+} = 4 \sqrt{P_{c}P_{s}}\; \Im \left[\frac{T \; e^{i\Omega\tau}\;(1+Re^{2i\Omega\tau})}{(Re^{2i\Omega\tau}-1)^{2}} \right]\;\frac{\delta\omega_{c}}{\Delta\nu_{\,\textnormal{FSR}}}.
\end{equation}
Under the assumption of high finesse, as the modulation frequency is much smaller than the FSR, the term below behaves as::
\begin{equation}
     \Im \left[\frac{T \; e^{i\Omega\tau}\;(1+Re^{2i\Omega\tau})}{(Re^{2i\Omega\tau}-1)^{2}} \right] \sim  \; -2\,\Im \left[\frac{1}{e^{2i\Omega\tau}-1} \right] \sim \frac{1}{\Omega\tau}.
\end{equation}
In the equation above, we performed a Laurent expansion around $\Omega\tau=0$. Substituting this expression into eq.~\ref{eq:ES1_lin1}, we obtain the error signal within its linear range:
\begin{equation}\label{eq:ES1_lin}
    \epsilon_{+} \simeq 8 \sqrt{P_{c}P_{s}} \;\frac{\delta\omega_{c}}{\Omega} = 16 \sqrt{P_{c}P_{s}} \;\frac{\delta f_{c}}{\Delta f} = D_{+} \; \delta f_{c},
\end{equation}

where we have written the the error signal in terms of the regular frequency $\delta f_{c}=\delta\omega_{c}/(2\pi)$ and defined the proportionality constant between $\epsilon_{+}$ and $\delta f_{c}$ as $D_{+}$, which is the optical gain of $L_{+}$ error signal relative to a variation of the laser frequency.

In the previous section we have shown that tuning the frequency splitting leads to a decrease in the power transmitted by the system. To simplify the notation, this effect has been intentionally disregarded in eq.~\ref{eq:single_cav}, where it is assumed that the maximum transmission coefficient remains 1 even for splittings larger than $\Delta f_{0}$. However, the consequence of having a lower transmitted power on the error signals is easy to incorporate. If, for example, the power is reduced by a factor of $\gamma$, the optical gain $D_+$ is reduced by the same amount.

As mentioned previously, this error signal utilizes the same symmetries as the PDH scheme. In fact, its mathematical derivation is equivalent to the ``Fast modulation near resonance'' case described in Black's work~\cite{Black2001}. The key properties of the PDH readout scheme are that the reflected signal is nearly zero at the carrier frequency, and the modulation sidebands are fully reflected by the cavity. This means that the ratio between the input power at the carrier frequency and that measured by the readout photodetector is at its maximum, while the same ratio for the sidebands is almost one.

In the proposed scheme, the photodetector in transmission fulfills the same role as the photodetector in reflection in the PDH scheme. The power reaching the readout photodetector is determined by the transmission coefficient of the coupled cavity $\xi(\omega)$. As mentioned earlier in eq.~\ref{eq:ES1_taylor}, this coefficient is at a minimum at the carrier frequency, $\xi(\omega_{c})\sim0$, while it is nearly one for the sidebands, $|\xi(\omega_{\pm})|\sim 1$. Therefore, the conditions for $L_{+}$ error signal are the same as those for the PDH error signal, making it formally equivalent to the latter. This is convenient as the PDH scheme is a well-performing technique in terms of noise couplings. According to~\cite{Black2001}, the PDH scheme is insensitive, at least to first order, to several noise sources: variation in the laser power, response of the photodiode used to measure the reflected signal, the modulation depth, the relative phase of the two signals going into the mixer, and the modulation frequency. 

For the sake of comparison with the PDH scheme, we rewrite $L_{+}$ error signal as
\begin{equation}
    \epsilon_{+} = 8 \sqrt{P_{c}P_{s}} \; \frac{T_1}{T_2}\;\frac{\delta f}{\Delta \nu_{\,\textnormal{LW}}}.
\end{equation}
As described in~\cite{Black2001}, the PDH error signal for a symmetric cavity with a length of $L$ and a transmission of $T=T_1/2$ through each mirror can be expressed as
\begin{equation}
    \epsilon_{\,\textnormal{PDH}} = 8 \sqrt{P_{c}P_{s}} \; \frac{\delta f}{\Delta \nu_{\,\textnormal{LW}}}
\end{equation}
Thus, assuming the same modulation depth, the ratio between the optical gain of the two error signals is
\begin{equation}
    \frac{\epsilon_{+}}{\epsilon_{\,\textnormal{PDH}}} = \frac{T_1}{T_2}
\end{equation}

\paragraph{Demodulating at $2\Omega$: $L_{-}$ error signal.}
Assuming demodulation at $2\Omega$, we isolate the $2\Omega$ terms in eq.~\ref{eq:analytics_Ptot} to obtain $L_{-}$ error signals:
\begin{equation}\label{eq:ES2_IQ}
  \epsilon_{-} = -2P_{s}\;\{\Re[G]\cos(\theta) + \Im[G]\sin(\theta)\}.
\end{equation}
As we are interested in the behavior of the error signal close to the operating point, we expand $\xi{(\omega)}$ in Taylor series around $\omega=\omega_{\pm}$, which gives
\begin{equation}
\xi(\omega_{\pm}) \simeq \mp 1 - i\frac{(1+R)T}{2(1-R)^{2}} \; \frac{\delta\omega_{m}}{\Delta\nu_{\,\textnormal{FSR}}},
\end{equation}
where $\delta \omega_{m}$ is the variation of the modulation frequency relative to $\Omega$.

As done in the previous section, the equation above can be simplified by assuming high finesse $\mathcal{F}\simeq \pi/(1-R)$, which also implies $(1+R)\sim2$. The previous equation becomes:
\begin{equation}
    \xi(\omega_{\pm}) \simeq \mp 1 - \frac{i}{\pi} \; \frac{\delta\omega_{m}}{\Delta\nu_{\,\textnormal{LW}}},
\end{equation}
Thus, $G$ can be expressed as:
\begin{equation}
    G = \xi(\omega_{+})\xi^{*}(\omega_{-}) \simeq -1 -\frac{2i}{\pi} \; \frac{\delta\omega_{m}}{\Delta\nu_{\,\textnormal{LW}}},
\end{equation}
where we only retained the terms linear in $\delta \omega_{m}$. By substituting this into eq.~\ref{eq:ES2_IQ}, we obtain the following for the $L_{-}$ error signal:
\begin{equation}
    \epsilon_{-} = 2P_{s}\;\left\{\cos(\theta) +\frac{2}{\pi} \; \frac{\delta\omega_{m}}{\Delta\nu_{\,\textnormal{LW}}}\sin(\theta)\right\}.
\end{equation}
\begin{figure}[htb]
    \centering
    \includegraphics{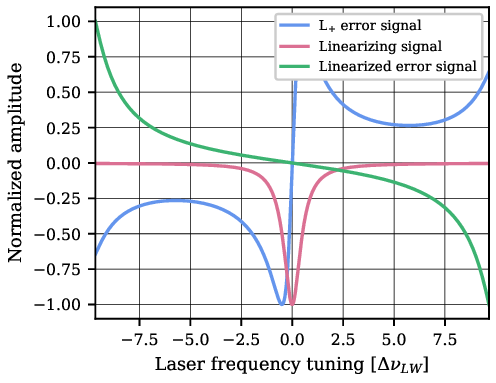}
    \caption{The linearization of the $L_{+}$ error signal. The standard $\epsilon_{+}$ error signal is shown in blue, while the linearizing signal, which corresponds to the quadrature component of the $L_{-}$ error signal, is represented in red. The green line represents the linearized error signal.}
    \label{figure:8}
\end{figure}
As the sine term in the previous equation contains all the information about the frequency change $\delta \omega_{m}$, we set the demodulation phase to $\theta=\pi/2$ to extract that term\footnote{With respect to a standard PDH scheme, the $2\Omega$ signal owns this term proportional to $\delta f_{m}$. This is due to the fact that the sidebands are transmitted to the photodetector as $\xi(\omega_{\pm})=\mp1$. In the PDH case, the reflection coefficient to the detector for the sidebands is -1, which would result in the cancellation of the $\delta f_{m}$ term.}. Within its linear region, the error signal is:
\begin{equation}\label{eq:ES2_lin}
    \epsilon_{-} = \frac{4P_{s}}{\pi} \; \frac{\delta\omega_{m}}{\Delta\nu_{\,\textnormal{LW}}} = 8P_{s} \frac{\delta f_{m}}{\Delta\nu_{\,\textnormal{LW}}} = D_{-} \; \delta f_{m}.
\end{equation}
As done in the previous section, we express the error signal in terms of regular frequency. $D_{-}$ is the optical gain of $L_{-}$ error signal relative to the variation of the modulator's frequency. Fig.~\ref{figure:6} presents a comparison between the output obtained from \finesse~and the analytical results derived in this section.

Applying the same considerations as for the $L_{+}$ error signal, when the frequency splitting is tuned to $\Delta f>\Delta f_0$, the optical gain $D_{-}$ is diminished by a factor of $\gamma$.

It is worth noting that the real component of the demodulated $2\Omega$ signal in (\ref{eq:ES2_IQ}) can be used to expand the linear region of the $L_{+}$ error signal. This is a useful feature that can make the process of bringing the system to the operating point simpler.

As evident in eqs.~\ref{eq:ES1_lin}, the linear region of $\epsilon_{+}$ error signal is noticeably narrow, and its characteristic length is set by the linewidth of the resonant peaks. The basic idea behind the linearization of an error signal is to divide it by another signal whose characteristic length is also the linewidth, thus expanding the linear region by a factor of roughly $\Delta\nu_{\,\textnormal{LW}}$. For the PDH scheme, this is achieved by dividing the error signal by the power transmitted through the cavity~\cite{Evans2002}. Furthermore, applications have shown that the same result can be achieved by utilizing the signal obtained by demodulating the photodiode in reflection at twice the modulation frequency~\cite{galaxies10060115}. 
In the context of coupled cavities, this can be done using the in-quadrature component of the $L_{-}$ error signal. When varying the laser frequency relative to the operating conditions, this is
\begin{multline}
    -2P_{s} \, \Re[G] = -2P_{s}|\xi(\omega_{\pm}+\delta\omega_{c})|^{2} \\
    = -2P_{s} \left|\frac{T\;e^{-i\delta\omega_c\tau}}{R\;e^{ -2i\delta\omega_c\tau}-1}\right|^2.
\end{multline}
The term $|\xi(\omega_{\pm}+\delta\omega_{c})|^{2}$ is  the Airy distribution, whose characteristic length is by definition the linewidth of the resonant peak. We define the linearized $L_{+}$ error signal as
\begin{equation}
    \Gamma_{+} = \frac{\epsilon_{+}}{-2P_{s} \, \Re[G]}.
\end{equation}
The linearization of the error signal is shown in Fig.~\ref{figure:8}.
%
%
\bibliography{biblio}
%
\end{document}